\documentclass[twocolumn,showpacs,preprintnumbers,amsmath,amssymb,eqsecnum]{revtex4}
\usepackage{graphicx}
\begin{document}
\preprint{IMAFF-RCA-03-03}
\title{Sub-Quantum Dark Energy}
\author{Pedro F. Gonz\'{a}lez-D\'{\i}az}
\affiliation{Centro de F\'{\i}sica ``Miguel A. Catal\'{a}n'', Instituto de
Matem\'{a}ticas y F\'{\i}sica Fundamental,\\ Consejo Superior de
Investigaciones Cient\'{\i}ficas, Serrano 121, 28006 Madrid (SPAIN).}
\date{\today}
\begin{abstract}
A procedure is considered which upgrades the Lagrangian
description of quantum relativistic particles to the Lagrangian of
a proper field theory in the case that the Klein-Gordon wave
equation is classically interpreted in terms of a relativistic
sub-quantum potential. We apply the resulting field theory to
cosmology and show that the relativistic version of the Bohm's
subquantum potential which can be associated with a homogeneous
and isotropic distribution of particles behaves like though it was
a cosmological constant responsible for the current accelerating
expansion of the universe, at least in the limit where the field
potential vanishes.
\end{abstract}


\maketitle

\section{Introduction}

With the advent of the increasingly compelling evidence for an
accelerating expansion of the universe [1-4], the traditional
problem of the cosmological constant [5], which disturbed
theoreticians a couple of decades ago, has become now even more
acute [6], as it has actually been replaced for the problem of the
nature of the so-called dark energy [7]. In scarcely five years a
rather impressive influx of papers have been published which tried
to shed some light at the question of what stiff is homogeneously
and isotropically pervading the universe up to nearly seventy
percent of its whole energy content and is able to currently make
it to evolve as though it was dominated by an anti-gravity regime
[8]. Several candidates have been claimed to make up this dark
energy. The first and most obvious option was - and still
continues being- a positive cosmological constant. The problem
with this interpretation is similar to the old
cosmological-constant problem: why the currently required value of
this quantity is so much smaller than that is predicted by
fundamental theories? [5]. Promoting the cosmological constant to
the status of a vacuum field having its own dynamics has raised a
rich plethora of possibilities which are described by means of a
slowly-varying scalar field generically denoted as either
quintessence models [9] or tracking models [10], depending on
whether the parameter $\omega$ entering the equation of state is
or is not a constant. Recent constraints on that equation of state
seem [11] nevertheless to point towards a value $\omega=-1$, which
corresponds to the cosmological constant, or even [11] to values
$\omega<-1$ which are associated with the so-called phantom
dark-energy models which predict a big rip in finite time [12].
Generalized Chaplygin gas models [13] and even further
generalizations from them [14] have also been recently claimed to
describe unified models of dark matter and energy. All of such
models have the interesting characteristic of describing by means
of a single entity- the gas- both the dark matter and dark energy
as given limiting cases. Finally, I would like to also mention the
so-called tachyon model [15] for dark energy which quite
remarkably construct a concept for dark energy from very
fundamental and simple relativistic principles which are promoted
to their field-theory counter-parts. Although there are many other
interpretations for dark energy, a review of which can be found in
Refs. [7,8], those mentioned so far look as being the most
promising.

In this paper we shall suggest a new interpretation for dark
energy which is essentially based on a generalization of the
tachyon model. Keeping in mind the idea that dark energy should
somehow reflect the otherwise unobservable existence of a
cosmological substance which will also have an essentially
quantum-mechanical nature, and promoting the so-called Bohm's
classical interpretation of quantum mechanics [16] to the status
of a field theory in a similar way to how it is made from
classical relativistic mechanics to finally produce the model of
tachyonic dark energy [15], we will thus be able to finally
propose simple "classical" field-theory models for dark energy
which does not necessarily depend on the existence of any
potential for the vacuum scalar field, and bring the imprint of
their truly quantum origin, formally in much the same way as
Bohm's classical interpretation of quantum mechanics does.

Starting with an approximate approach, we develop some fundamental
aspects of the field theory resulting from upgrading the
"classically" interpreted quantum particle properties to field
variables, which we also connect to current flat cosmology. Among
the different limiting situations that we shall consider from that
field theory, we single out one with vanishing field potential as
the most interesting interpretation for dark energy in this paper:
It is the sub-quantum potential associated with the particles
which should be interpreted as being the cosmological constant
responsable for dark energy.

The paper can be outlined as follows. In Sec. II we shall discuss
a formalism based on applying some approximations to the
relativistic energy-momentum relation with a sub-quantum
potential. Fundamental aspects of the field theory are discussed
in Sec. III where a recipe is given to compute the field
potential. Sec. IV contains the description of a cosmological
model which is based on coupling the full field Lagrangian to
Hilbert-Einstein gravity. We summarize and conclude in Sec. V,
where we also add some relevant discussions and comments.

\section{An approximate model}

From the real part of the Klein-Gordon wave equation applied to a
quasi-classical wave function $R\exp(iS/\hbar)$, where the
probability amplitude $R$ ($P=|R|^2$) and the action $S$ are real
functions of the relativistic coordinates, if the classical energy
$E=\partial S/\partial t$ and momentum $p=\nabla S$ are defined,
one can write [15]
\begin{equation}
E^2-p^2+V_{SQ}^2=m_0^2,
\end{equation}
where $m_0$ is the rest mass of the involved particle and $V_{SQ}$
is a relativistic sub-quantum potential,
\begin{equation}
V_{SQ}^2= \frac{\hbar^2}{R}\left(\nabla^2 R-\frac{\partial^2
R}{\partial t^2}\right) ,
\end{equation}
which should be interpreted according to the Bohm's idea [16] as
the hidden sub-quantum potential that accounts for precisely
defined unobservable relativistic variables whose effects would
physically manifest in terms of the indeterministic behaviour
shown by the given particles. From Eq. (2.1) it immediately
follows that $p=\sqrt{E^2+V_{SQ}^2-m_0^2}$. Thus, since
classically $p=\partial L/\partial[\dot{q(t)}]$ (with $L$ being
the Lagrangian of the system and $q$ the spatial coordinates which
depends only on time $t$, $q\equiv q(t)$), we have for the
Lagrangian
\begin{equation}
L=\int d\dot{q}p=\int dv\sqrt{\frac{m_0^2}{1-v^2}+M^2} ,
\end{equation}
in which $v=\dot{q}$ and $M^2=V_{SQ}^2-m_0^2$. The latter quantity
is by no means ensured to be positive definite in the general
theory. In fact, one could easily consider rest masses that would
exceed the sub-quantum potential. Actually, if the particle is
assumed to move locally according to some causal law, then the
classical expression $E=\partial S/\partial t$ and $p=\nabla S$,
where $S$ is the classical relativistic action, are locally
satisfied. Hence, we can average Eq. (2.1) with a probability
weighting function $P(x,t)=|R(x,t)|^2$, so that
\begin{eqnarray}
&&\int\int\int dx^3 P(x,t)\left(E^2-p^2+V_{SQ}^2\right)\nonumber\\
&&=\langle E^2\rangle_{\rm av}-\langle p^2\rangle_{\rm av}
+\langle V_{SQ}^2\rangle_{\rm av}=m_0^2 ,\nonumber
\end{eqnarray}
with the averaged quantities coinciding with the corresponding
classical quantities and $\langle V_{SQ}^2\rangle_{\rm av}$ being
a constant. Now, the velocity of the particle can be defined to be
\[\langle v\rangle_{\rm av} =\frac{\langle p^2\rangle_{\rm
av}^{1/2}}{\left(\langle p^2\rangle_{\rm av} -\langle
M^2\rangle_{\rm av}\right)^{1/2}} ,\] in which $\langle
M^2\rangle_{\rm av}=\langle V_{SQ}^2\rangle_{\rm av}-m_0^2$. Thus,
in the general particle theory, "effective" slower than light,
faster than light and light-like particles can be defined to
respectively satisfy $m_0 > \langle V_{SQ}^2\rangle_{\rm av}$,
$m_0 < \langle V_{SQ}^2\rangle_{\rm av}$ and $m_0 = \langle
V_{SQ}^2\rangle_{\rm av}$, so implying different values and signs
for the quantity $\langle M^2\rangle_{\rm av}$, and hence $M^2$.

In the classical limit $\hbar\rightarrow 0$, $V_{SQ}\rightarrow
0$, and hence we are left with just the classical relativistic
Lagrangian for a particle with rest mass $m_0$. As shown by Bagla,
Jassal and Padmanabhan [15], promoting the quantities entering
this simple Lagrangian to their field-theory counter-parts allows
us to get a cosmological model with tachyonic dark energy. In what
follows of this section we shall explore the question of what kind
of cosmological models can be derived if we apply an
upgrading-to-field procedure starting with Lagrangian (2.3). For
such an upgrading formalism we shall use throughout the paper the
one employed by Padmanabhan {\it et al.} [14,17,18], except for
the harmless presence of the constant sub-quantum potential. Our
procedure is thus based on upgrading the coordinate $q(t)$ to a
field $\phi$ which, by relativistic invariance, will depend on
both space and time while $\dot{q}^2$ is replaced for the quantity
$\partial_i\phi\partial^i\phi$. This makes also possible to regard
the mass parameter $m_0$ as a potential function of field $\phi$,
thereby obtaining a given field theoretic Lagrangian. The
Hamiltonian structure of the resulting theory will be
algebraically similar to that of special relativity equipped with
a sub-quantum potential term. The theory will in this way allow
for solutions depending on both space and time, with finite
momentum and energy densities analogous to some descriptions
arising in string theory. For more details on that formalism and
its motivation I address the reader to Refs. [14,17,18].

Two approximate limiting situations will be considered in this
section, starting with particle-theory cases where $m_0\leq
V_{SQ}$ which ensure that $M^2>0$. First of all, we shall look at
the case of most cosmological interest which corresponds to the
limit of small values of the rest mass, $m_0\rightarrow 0$, for
which the Lagrangian becomes
\[L\simeq\sqrt{V_{SQ}^2-m_0^2}\int dv\left(1+
\frac{m_0^2}{2\left(V_{SQ}^2-m_0^2\right)(1-v^2)}\right) \]
\begin{equation}
=\sqrt{V_{SQ}^2-m_0^2}v+
\frac{m_0^2}{\sqrt{V_{SQ}^2-m_0^2}}\ln\left[
\left(\frac{1+v}{1-v}\right)^{1/4}\right].
\end{equation}
This Lagrangian is positive definite whenever $V_{SQ}>0$. For
nonzero values of the sub-quantum potential, we can have physical
systems with nonzero Lagrangian even for the massless case where
$v=1$ and $m_0=0$ simultaneously. This is made possible since the
existence of the sub-quantum potential allows us to consider an
"effective" rest mass given by $M\equiv\sqrt{V_{SQ}^2-m_0^2}$. On
the other hand, since the sub-quantum potential $V_{SQ}$ can take
on both positive and negative values, the associated field theory
can lead to positive or negative pressure, respectively. Choosing
$V_{SQ}<0$ and hence $L<0$, in the massless case $m_0=0$, $v=1$,
we have
\begin{equation}
L=-|V_{SQ}|.
\end{equation}
Generalizing to a field theory in the general case $m_0\neq 0$,
$v<1$ requires the upgrading $q(t)\rightarrow \phi$, a field which
will thereby depend on both space and time, $\phi(r,t)$, replacing
$v^2\equiv\dot{q}^2$ for $\partial_i\phi\partial^i\phi$ and the
rest mass $m_0$ for a generic potential $V(\phi)$). In the extreme
massless case however the Lagrangian (2.5) does not contain any
quantity which can be upgraded to depend on $\phi$, so that the
Lagrangian for the field theory in the massless case is no longer
zero, but it is also given by Eq. (2.5).

We shall regard here Lagrangian (2.5) as containing all the
cosmological information that corresponds to a universe whose dark
energy is given by a positive cosmological constant, provided the
field $\phi$ is homogeneously and isotropically distributed. This
can be accomplished if e.g. the sub-quantum potential is
interpreted as that potential associated to the hidden dynamics of
the particles which are homogeneously and isotropically
distributed in the universe. Assuming next a perfect fluid form
for the equation of state of the cosmic field $\phi$, i.e.
introducing a stress-energy tensor
\begin{equation}
T_k^i = (\rho+P)u^i u_k - p\delta_k^i,
\end{equation}
where the energy density $\rho$ and the pressure $p$ that
correspond to Lagrangian (2.5) are given by
\begin{equation}
\rho=|V_{SQ}|,\;\;\; p=-|V_{SQ}| ,
\end{equation}
and the 4-velocity is
\begin{equation}
u_k= \frac{\partial_k\phi}{\sqrt{\partial_i\phi\partial^i\phi}} .
\end{equation}
From Eqs. (2.7) and the conservation equation for cosmic energy,
$d\rho=-3(\rho+p)da/a$, it again follows that
$\rho=\kappa^2=|V_{SQ}|={\rm const.}$, so that the resulting
Friedmann equation, $\dot{a}=\kappa a/m_P$ ($m_P$ being the Planck
mass), yields the expected solution for the scale factor
$a=a_0\exp\left[\kappa(t-t_0)/m_P\right]$. Eqs. (2.7) immediately
lead moreover to a characteristic parameter for the perfect fluid
state equation which turns out to be constant and given by
$\omega=P/\rho=-1$. We can conclude therefore that if $m_0=0$,
$v=1$ (i.e. $V(\phi)=0$ and $\partial_i\phi\partial^i\phi=1$ in
the field theory) and $V_{SQ}<0$, the hidden dynamics of particles
or fields makes to appear a sub-quantum potential inducing the
presence of a pure cosmological constant given by
$\Lambda=\kappa=\sqrt{V_{SQ}}$. In case that the rest mass is
$m_0\neq 0$ and very small, there would be a nonzero field-theory
potential $V(\phi)\rightarrow m_0$ and the sub-quantum medium
would correspond to a cosmic dark energy which would somehow
behave like some sort of "tracking" quintessential field [10]. In
fact, for in such a case we had for negative $V_{SQ}$ and small
but nonzero $m_0$,
\begin{equation}
L=p=-|M|\sqrt{\partial_i\phi\partial^i\phi}-
\frac{V(\phi)^2}{4|M|}\ln\left(\frac{1+
\sqrt{\partial_i\phi\partial^i\phi}}{1-\sqrt{\partial_i\phi\partial^i\phi}}\right),
\end{equation}
with $M$ being now given by $M\equiv
M[V(\phi)]=-\sqrt{V_{SQ}^2-V(\phi)^2}$, where $V(\phi)$ is again
generally defined through the upgrading procedure, i.e.
$m_0\rightarrow V(\phi)$. The pressure $p$ is then a definite
negative quantity such that $\partial_i\phi\partial^i\phi
<2V(\phi)$ only if $\partial_i\phi\partial^i\phi$ is sufficiently
smaller than $\left(\partial_i\phi\partial^i\phi\right)_c$, with
\[\frac{\sqrt{\left(\partial_i\phi\partial^i\phi\right)_c}}{1-
\left(\partial_i\phi\partial^i\phi\right)_c}= \ln\left[\frac{1+
\sqrt{\left(\partial_i\phi\partial^i\phi\right)_c}}{\sqrt{1-
\left(\partial_i\phi\partial^i\phi\right)_c}}\right] . \]

The energy density which together with the pressure $p$ enters the
equation of state $p=\omega(\phi)\rho$ would then read
\begin{equation}
\rho=-\frac{V(\phi)^2}{2|M(\phi)|}\left[\frac{\sqrt{\partial_i\phi
\partial^i\phi}}{1- \partial_i\phi\partial^i\phi}-\ln\left(\frac{1+ \sqrt{\partial_i\phi\partial^i\phi}}{\sqrt{1-
\partial_i\phi\partial^i\phi}}\right)\right] .
\end{equation}
We then note that for the considered range of the kinetic term, we
always can in fact choose a range for the parameter entering the
equation of state which satisfies $0\geq\omega(\phi)\geq -1$.

In the limit that the rest mass and the sub-quantum potential take
on very similar values, which is the second situation we shall
briefly consider, the Lagrangian can be approximated to
\begin{equation}
L\simeq m_0\int\frac{dv}{\sqrt{1-
v^2}}=\frac{1}{2}m_0\ln\left(\frac{1-
\sqrt{1-v^2}}{1+\sqrt{1-v^2}}\right)^{1/2} .
\end{equation}
Such a Lagrangian is negative definite and, if we upgrade the
quantities involved in it so that they become field-theory
variables, $m_0\rightarrow V(\phi)$, with $V(\phi)$ a classical
potential for the scalar field $\phi$, and
$v^2\rightarrow\partial_i\phi\partial^i\phi$, it would correspond
to a negative pressure
\begin{equation}
p=\frac{1}{2}V(\phi)\ln\left(\frac{1-\sqrt{1-
\partial_i\phi\partial^i\phi}}{1+\sqrt{1-\partial_i\phi\partial^i\phi}}\right),
\end{equation}
which is definite negative, and a positive energy density
\begin{equation}
\rho=\frac{V(\phi)}{\sqrt{1-\partial_i\phi\partial^i\phi}}-p .
\end{equation}
Thus, for a perfect fluid equation of state $p=\omega(\phi)\rho$,
this would again be somehow analogous to a tracking
quintessence-like field. Anyway, the simplest and most interesting
situation we have dealt with in this section corresponds to the
massless case where the consideration of the particles in terms of
the relativistic version of the classical Bohm's theory is enough
to promote their corresponding sub-quantum potential to the same
status as that of a cosmological constant. Therefore, if we adopt
the interpretation considered in this letter, dark energy appears
to at least partly correspond to the overall work which is done by
all particles along their hidden trajectories associated through
the Bohm's interpretation with the essential quantum indeterminacy
of the observable matter in the universe.

\section{The field theory}

In order to construct a field theory starting with the classical
relation $E^2-p^2+V_{SQ}^2=m_0^2$, one should first integrate Eq.
(2.3) to yield the Lagrangian in closed form
\begin{equation}
L=-m_0 E\left(x,k\right) ,
\end{equation}
where $E(x,k)$ is the elliptic integral of the second kind [16],
with
\begin{equation}
x=\arcsin\sqrt{1-v^2} ,\;\;\; k=\sqrt{1-\frac{V_{SQ}^2}{m_0^2}} .
\end{equation}
The Lagrangian (3.1) describes a relativistic particle with a (one
dimensional) position $q(t)$ and a mass $m_0$ whose local motion
is causally disturbed by the presence of a sub-quantum potential
$V_{SQ}$. One can now proceed to upgrading [14] the particle
theory into the field theory by promoting $m_0\rightarrow V(\phi)$
and $v^2\rightarrow\partial^i\phi\partial_i\phi$ in Lagrangian
(3.1). Thus, we obtain
\begin{equation}
L=-V(\phi)E\left(x(\phi),k(\phi)\right) ,
\end{equation}
in which
\begin{equation}
x(\phi)=\arcsin\sqrt{1-\partial^i\phi\partial_i\phi} ,\;\;\,
k(\phi)=\sqrt{1-\frac{V_{SQ}^2}{V(\phi)^2}} .
\end{equation}
We shall restrict ourselves in Secs. III and IV to consider the
field theoretic analogous to "effective" slower than light
relativistic particles filling a flat universe with the
conventional scalar field $\phi$ having the potential $V(\phi)$ as
a source. The evolution of that universe will be assumed to be
specified so that the scale factor $a(t)$ and its time derivatives
$H(t)\equiv(\dot{a}/a)$,... are all known functions of time $t$.
For a Friedmann-Robertson-Walker (FRW) universe $\phi(t,{\bf
x})=\phi(t)$. Thus, our problem will be to determine $\phi$ and
hence the potential for the scalar field $V(\phi)$ as given
functions of the specified cosmic parameters $H(t)$ and
$\dot{H}(t)$.

The Friedmann equations are
\begin{equation}
H^2=\frac{8\pi G}{3}\rho_T ,\;\; \frac{\ddot{a}}{a}=-\frac{4\pi
G}{3}\left(\rho_T+3p_T\right) ,
\end{equation}
with $\rho_T=\rho_{ob}+\rho_{\phi}$ the energy density for
observable matter plus dark energy, and $p_T$ the corresponding
sum of pressures. For a theory which generalizes the
non-relativistic description of a single particle, that is for a
typical quintessence theory, the field $\phi(t)$ is defined by the
conventional expressions
\begin{equation}
\rho_{\phi}=\frac{1}{2}\dot{\phi}^2 +V(\phi) ,\;\;
p_{\phi}=\frac{1}{2}\dot{\phi}^2 -V(\phi) .
\end{equation}

For the FRW universe the Lagrangian generalizing the relativistic
particle Lagrangian plus a sub-quantum potential reduces to be
given by Eq. (3.3) with
\begin{equation}
x(\dot{\phi})=\arcsin\sqrt{1-\dot{\phi}^2} ,\;\;\,
k(\phi)=\sqrt{1-\frac{V_{SQ}^2}{V(\phi(t))^2}} .
\end{equation}
In the limit $V_{SQ}\rightarrow 0$, that Lagrangian becomes the
Lagrangian for the tachyon model [14,17,18], i.e.
$L=-V(\phi)\sqrt{1 -\dot{\phi}^2}$, as it should be expected. The
pressure and energy density for our field model to be used in Eq.
(3.5) are no longer given by Eqs. (3.6), but the new
"relativistic" expressions
\begin{equation}
p_{\phi}=-V(\phi)E(x,k)
\end{equation}
\begin{equation}
\rho_{\phi}= \frac{V(\phi)\sqrt{1- \frac{\Delta
V^2(1-\dot{\phi}^2)}{V(\phi)^2}}\dot{\phi}}{\sqrt{1-
\dot{\phi}^2}}+V(\phi)E(x,k) ,
\end{equation}
where $\Delta V^2=V(\phi)^2-V_{SQ}^2$. Notice that (i) in the
limit $V_{SQ}\rightarrow 0$ $\rho_{\phi}\rightarrow
V(\phi)/\sqrt{1-\dot{\phi}^2}$ and $p_{\phi}\rightarrow
V(\phi)\sqrt{1-\dot{\phi}^2}$, i.e. the expressions that
respectively correspond to the energy density and pressure in the
tachyon model [14,17,18]; and (ii) the state-equation parameter
$\omega(t)=p_{\phi}/\rho_{\phi}$ derived from Eqs. (3.8) and (3.9)
is generally larger than -1 and is therefore associated with a
dark energy content which does not match a cosmological constant.
In any case, for a source with parameter
$\omega(t)=p_{\phi}/\rho_{\phi}$, we must always have [17]
\begin{equation}
\frac{\dot{\rho}_{\phi}}{\rho_{\phi}}=
-3H(1+\omega)=\frac{2\dot{H}}{H} .
\end{equation}

Combining Eq. (3.10) with the expression for $\omega(t)$, we
obtain
\begin{equation}
\frac{\sqrt{1-\frac{\Delta V^2(1-
\dot{\phi}^2)}{V(\phi)^2}}\dot{\phi}}{\sqrt{1-\frac{\Delta V^2(1-
\dot{\phi}^2)}{V(\phi)^2}}\dot{\phi}+E(x,k)\sqrt{1-\dot{\phi}^2}}
=-\frac{2\dot{H}}{3H^2} .
\end{equation}
On the other hand, multiplying Eqs. (3.8) and (3.9) and using Eq.
(3.10) and the Friedmann equation, one can derive
\begin{equation}
V(\phi)= \frac{\frac{3H^2}{8\pi G}\left(1+
\frac{2\dot{H}}{3H^2}\right)^{1/2}}{\left\{\left[\frac{\sqrt{1-\frac{\Delta
V^2(1-
\dot{\phi}^2)}{V(\phi)^2}}\dot{\phi}}{\sqrt{1-\dot{\phi}^2}}
+E(x,k)\right]E(x,k)\right\}^{1/2}} ,
\end{equation}
where
\begin{equation}
E(x,k)=-\sqrt{\frac{1-\frac{\Delta V^2(1-
\dot{\phi}^2)}{V(\phi)^2}}{1-\dot{\phi}^2}}\dot{\phi}\left(1+
\frac{3H^2}{2\dot{H}}\right) .
\end{equation}

Eqs. (3.11) and (3.12) would solve the problem posed initially,
that is the problem of obtaining the field potential in terms of
the scalar field, $V(\phi)$, and the scalar field in terms of
time, $\phi(t)$. In fact, from the above equations we can obtain a
simpler expression for $V(\dot{\phi})$ which is given by
\begin{equation}
V(\dot{\phi})= -\left[\left(\frac{2\dot{H}}{8\pi G}\right)^2
-\dot{\phi}^2
V_{SQ}^2\right]\frac{\sqrt{1-\dot{\phi}^2}}{\dot{\phi}^2} .
\end{equation}
Now, by inserting $V(\dot{\phi})$ into Eq. (3.11) and integrating
the resulting expression over time $t$ one would attain an
expression for $\phi(t)$. Re-expressing then $\phi(t)$ as a
function of $\dot{\phi}(t)$ and the given scale factor $a(t)$ an
expression for $V(\phi)$, compatible with the original hypothesis,
could finally be obtained. In principle, given any scale factor
$a(t)$, one can then obtain $V(t)$ and $\phi(t)$, and hence the
potential $V(\phi)$ by using Eqs. (3.11), (3.12) and (3.14).
However, these equations are very complicated and cannot be
inmediately solved except for certain limiting approximated cases.
In what follows of this section, we shall consider two of such
limiting cases where these equations can be easily solved.
Firstly, if we let $V_{SQ}\rightarrow 0$, we inmediately recover
the tachyon-model expression which was already dealt with by
Padmanabhan [17]. If, on the other hand, one lets
$V(\phi)^2\rightarrow V_{SQ}^2$, then we can derive the
approximate expressions
\[\frac{\dot{\phi}}{\dot{\phi} +\sqrt{1-\dot{\phi}^2}\arcsin\sqrt{1-
\dot{\phi}^2}}\simeq-\frac{2\dot{H}}{3H^2} \]
\[V\simeq \frac{\frac{3H^2}{8\pi G}\left(1
+\frac{2\dot{H}}{3H^2}\right)^{1/2}\left(1-
\dot{\phi}^2\right)^{1/4}}{\sqrt{\left(\dot{\phi}
+\sqrt{1-\dot{\phi}^2}\arcsin\sqrt{1-
\dot{\phi}^2}\right)\arcsin\sqrt{1-\dot{\phi}^2}}} .\] Restricting
ourselves then to the late-time regime of slowly-varying field,
$\dot{\phi}\rightarrow 0$, and nearly constant $H$, we finally get
\begin{equation}
\phi\simeq -\pi\int\frac{\dot{H}dt}{3H^2} ,\;\; V\simeq
\frac{3H^2}{4\pi^2 G}\left(1+
\frac{2\dot{H}}{3H^2}\right)\rightarrow V_{SQ} .
\end{equation}
These values of $\phi$ and $V\simeq V_{SQ}$ are very similar
indeed to those of the tachyon model [14,17,18]. They are in fact
approximately the same unless by a factor $\pi/2$ in $\phi$ and by
a factor $2/\pi$ in $V(t)$. If we would finally consider a late
universe with power law expansion $a=t^n$, then it turned out
$\phi\simeq\pi t/3n +\phi_0$, which resulted in a potential
$V(\phi)\simeq 3/[12G(\phi-\phi_0)^2]$.

\section{A cosmological model}

Let us consider in this section the observationally most favored
case of a spatially flat universe with Friedmann equations given
by Eqs. (3.5). For our scalar field we have an energy density and
a pressure as given in Eqs. (3.8) and (3.9), respectively. Writing
the stress tensor in the perfect-fluid form shown by Eq. (2.6) and
four-velocity $u_k$ as expressed in Eq. (2.8), we now can uncover
that the stress tensor for field $\phi$ may be regarded as being
the sum of a pressureless dust component plus a vacuum
cosmological-constant-like component. This can most clearly be
seen by breaking up the energy density $\rho_{\phi}$ and pressure
$p_{\phi}$ as [14,18]: $\rho_{\phi}=\rho_v +\rho_{DM}$,
$p_{\phi}=p_v +p_{DM}$, with
\begin{equation}
\rho_v=V(\phi)E(x,k),\;\;
\rho_{DM}=\frac{V(\phi)}{\sqrt{1-\dot{\phi}^2}}\sqrt{1-
\frac{\Delta V^2 (1-\dot{\phi}^2)}{V(\phi)^2}}
\end{equation}
\begin{equation}
p_v=-V(\phi)E(x,k),\;\;\; p_{DM}=0 ,
\end{equation}
where we note that $\omega_v=p_v/\rho_v=-1$, so that dark energy
behaves like a cosmological constant.

The present description resembles that of generalized
Chaplygin-gas models [12], which likewise describe could dark
matter and pure dark energy as the extreme limiting cases for a
single field at the highest and smallest densities, respectively.
In the present case, the choice in Eqs. (4.1) and (4.2) allows us
to write for early time
\begin{equation}
\lim_{t\rightarrow 0}\left[V(\phi)E(x,k)\right]\rightarrow 0 ,
\end{equation}
and therefore $\dot{\phi}\rightarrow\pm 1$ as $t\rightarrow 0$. On
the other hand, for large $t$,
\begin{equation}
\lim_{t\rightarrow\infty}\left[\frac{V(\phi)}{\sqrt{1-
\dot{\phi}^2}}\sqrt{1- \frac{\Delta V^2(1-
\dot{\phi}^2)}{V(\phi)^2}}\right]\rightarrow 0 ,
\end{equation}
which in turn implies that for $V(\phi)\neq 0$, $\Delta V^2(1-
\dot{\phi}^2)/V(\phi)^2\rightarrow 1$; that is, if for example
$V(\phi)^2 >> V_{SQ}$, then $\dot{\phi}\rightarrow 0$, so matching
the classical, slowly-varying behaviour of a scalar field that
characterizes quintessential models.

On the other hand, the action that couples the scalar field $\phi$
and the given observable matter fields to Hilbert-Einstein gravity
can be written as
\begin{equation}
S=\int d^4 x\sqrt{-g}\left[\frac{R}{16\pi G}-
V(\phi)E(x,k)+L_{ob}\right] ,
\end{equation}
where $L_{ob}$ is the Lagrangian for the observable matter field
we should add to dark matter and energy in order to represent a
realistic universe. The latter non-gravitational components, i.e.
dark matter and dark energy, are both unitarily described by just
the scalar field $\phi$. We can now derive the equation of motion
for such a field in our cosmological context to be
\begin{widetext}
\begin{eqnarray}
&&\dot{\phi}\ddot{\phi}= -(1-\dot{\phi}^2)\sqrt{1-(1-
\dot{\phi}^2)\left(1- \frac{V_{SQ}^2}{V(\phi)^2}\right)}
\left[3H\sqrt{1 -(1-\dot{\phi}^2)\left(1-
\frac{V_{SQ}^2}{V(\phi)^2}\right)}\right.\nonumber\\
&&\left.+\sqrt{1-
\dot{\phi}^2}\left(\frac{dV(\phi)}{V(\phi)d\phi}\right)
\left(\frac{E(x,k)V(\phi)^2
-F(x,k)V_{SQ}^2}{V(\phi)^2-V_{SQ}^2}\right)\right] ,
\end{eqnarray}
\end{widetext}
where
\begin{equation}
F(x,k)=F\left(\arcsin\sqrt{1-\dot{\phi}^2},\sqrt{1-
\frac{V_{SQ}^2}{V(\phi)^2}}\right)
\end{equation}
is the elliptic integral of the first kind [16], with the same
argument $x$ and parameter $k$ as the integral $E(x,k)$.

Inspection of Eq. (4.6) leads us to notice that, such as it
happens in the tachyon theory [14,17,18], the change of
$\dot{\phi}$ goes to zero as it approaches $\pm 1$ or, what is
specific of the present model, $\pm
V_{SQ}/[V(\phi)\sqrt{V_{SQ}^2/V(\phi)^2 -1}]$. In these cases the
equation of state for the field $\phi$ is dustlike. Thus, at any
stage if the field behaves like dust, it continues to do so a long
time. Such a behaviour persisted for a duration that depends on
the closeness of $\dot{\phi}$ to $\pm 1$ or $\pm
V_{SQ}/[V(\phi)\sqrt{V_{SQ}^2/V(\phi)^2 -1}]$. Detailed behaviour
of the field evolution would depend on the shape of the used
potential. On the other hand, the change of $\dot{\phi}$ goes to
infinity as it approached zero. In that case, it can be checked
that the energy density $\rho_v=-p_{v}$ should be given in terms
of the complete elliptic integral of the second kind [16],
$E\left(k=\sqrt{1-V_{SQ}^2/V(\phi)^2}\right)$, while $\rho_{DM}$
is given by $V_{SQ}$.

Even for the simplest possible potentials, the solutions to Eq.
(4.6) and the associated Friedmann equation is rather a cumbersome
task. Numerical calculations are left for a future work.

\section{Summary and discussion}

In this paper we have developed the essential, preliminary aspects
of a new field theory and its application to cosmology. It is
based on the novel idea that vacuum fields can correspond to an
up-grading of the Lagrangian of the quantum relativistic
description of a single particle, as interpreted in terms of the
"classical" sub-quantum potential, to a scalar field theory. Some
rather heuristic arguments were first advanced that led to the
interpretation that the field-theory Lagrangian approximately
becomes the sub-quantum potential in the limit of small values of
the field potential and largest values of the kinetic term. In
what follows we shall comments on this reduction process from
which we can deduce that the equation of state for dark energy at
that limit is $p=-\rho$; i.e. the equation of state for a pure
cosmological constant whose dynamics is fully hidden. A brief
classical description of the full field theory is then carried
out. This leads to formulate the way in which the potential for
the scalar field can be obtained in a cosmological context where
the universe is spatially flat. Some cosmic predictions are
discussed afterward basing on the formulation of the equation of
motion for the scalar field.

We shall now consider a couple of limiting cases from the full
field theory developed in Secs. III and IV. We first derive the
approximate form of the Lagrangian when we let $V(\phi)\rightarrow
0$, $\dot{\phi}\rightarrow\pm 1$, keeping $V_{SQ}$ nonzero. This
case was already considered starting directly from an approximate
expression for the classical momentum by using heuristic
arguments. We shall check that the results obtained in Sec. II are
again obtained by using Eq. (3.1). In fact, if
$V_{SQ}^2>>V(\phi)^2$, the Lagrangian in Eq. (3.1) can be written
as
\begin{widetext}
\begin{equation}
L=-V_{SQ}\left[E\left(\frac{V_{SQ}}{V(\phi)}x,\kappa\right)
+\kappa{\rm sn}\left(\sqrt{1-\dot{\phi}^2}\frac{V_{SQ}}{V(\phi)},
\kappa\right){\rm
cd}\left(\sqrt{1-\dot{\phi}^2}\frac{V_{SQ}}{V(\phi)},
\kappa\right)\right],
\end{equation}
\end{widetext}
where
\begin{equation}
\kappa=\sqrt{1-\frac{V(\phi)^2}{V_{SQ}^2}}
\end{equation}
and ${\rm sn}$ and ${\rm cd}$ are elliptic functions [16]. Now, in
the limit $\sqrt{1-\dot{\phi}^2}\rightarrow 0$ and
$V(\phi)\rightarrow 0$ the Lagrangian (5.1) reduces to
\begin{equation}
L\simeq -V_{SQ}\left[\sin\left(\frac{\sqrt{1-
\dot{\phi}^2}V_{SQ}}{V(\phi)}\right)-\tanh\left(\frac{\sqrt{1-
\dot{\phi}^2}V_{SQ}}{V(\phi)}\right)\right],
\end{equation}
which can be written as $L\sim -V_{SQ}$ since $\sqrt{1-
\dot{\phi}^2}/V(\phi)$ is an indeterminate quantity. Consistency
of this and the results to follow is ensured by checking that, in
fact, as one lets $V_{SQ}\rightarrow 0$ the Lagrangian of the
tachyon theory [14,17,18], $L=-V(\phi)\sqrt{1-\dot{\phi}^2}$, is
recovered from Eq. (3.1). The result in Eq. (5.2) can again be
derived in the limit where $V(\phi)\rightarrow V_{SQ}$; that is
\begin{equation}
L\simeq -V_{SQ}\arcsin\sqrt{1-\dot{\phi}^2} ,
\end{equation}
if one considers a slowly-varying field such that
$\dot{\phi}\rightarrow 0$. Thus, we again get $L\simeq-V_{SQ}$ and
hence once again $p\simeq -V_{SQ}$, $\rho\simeq V_{SQ}$, as
corresponding to a cosmological constant with state equation
parameter $\omega=-1$. If, instead of $\dot{\phi}\rightarrow 0$,
we had taken $\dot{\phi}\rightarrow \pm 1$, then a new field
theory at all similar to tachyon theory but with the potential
$V(\phi)$ replaced for the sub-quantum potential $V_{SQ}$ had been
obtained.

Consider finally the hypothetical case where $\dot{\phi}^2 > 1$.
If we then rotate the field-theory potential $V(\phi)$ to purely
imaginary values so that $V(\phi)\rightarrow iW(\phi)$, then the
Lagrangian would take the form
\begin{eqnarray}
&&L(\phi)=-W(\phi)\left[E(y,\gamma)-F(y,\gamma)\right.\nonumber\\
&&\left.-\sqrt{\dot{\phi}^2-1}\left(1 +\frac{(\dot{\phi}^2-
1)V_{SQ}^2}{\dot{\phi}^2 W(\phi)^2}\right)\right] ,
\end{eqnarray}
with
\begin{equation}
y=\arctan\sqrt{\dot{\phi}^2-1}, \;\; \gamma=\frac{V_{SQ}}{V(\phi)}
\end{equation}
and $F(y,\gamma)$ again being the elliptic integral of the first
kind. Clearly, in the limit $V_{SQ}\rightarrow 0$, we get what one
could denote as a tachyon field theory from tachyonic particles
for which
\begin{equation}
L\simeq W(\phi)\sqrt{\dot{\phi}^2-1} ,
\end{equation}
and hence
\begin{equation}
p\simeq W(\phi)\sqrt{\dot{\phi}^2-1}, \;\;
\rho\simeq\frac{W(\phi)}{\sqrt{\dot{\phi}^2-1}} .
\end{equation}
Note that in this case the pressure associated with the scalar
field is positive as that field must now be a
superluminarly-varying field and therefore
$\dot{\phi}^2/2>W(\phi)$.

Before closing up, I will add some comments which may be helpful
for a better understanding of some points of this paper. First of
all, we ought to notice that, besides the limiting case where
$v\rightarrow\pm 1$ and $m_0\rightarrow 0$, one should apply the
Bohm's interpretation and the upgrading procedure to vacuum
massive particles. In that case, too, the existence of a positive
pressure making the universe viable may be ensured by providing in
a quite natural way the observable Lagrangian $L_{ob}$ in action
(4.5) with a suitable equation of state. A mutual balance between
such a positive pressure and the negative pressure $p_{\phi}$
would naturally arises as a consequence from the comparison
between the equations of motion for the observable matter fields
and Eqs. (4.1), (4.2) and (4.6). What one would conventionally
expect from that comparison is that, after passing through a
pressureless regime with $\phi$-dark matter at the earliest
universe, this would become dominated first by a positive-pressure
regime with observable energy and then by the negative-pressure
late dark-energy $\phi$-regime, after reaching the coincidence
time [19].

In principle, there are two possible interpretations for field
$\phi$. On the one hand, $\phi$ could be viewed as the "classical"
field which arises from the hidden dynamics that led to the
apparent quantum behaviour of all particles and cosmic fields,
including CMB. If such an interpretation is accepted, then there
would not be anything like dark energy. On the other hand, field
$\phi$ could be also regarded as making up the real stiff for
existing dark energy. Although the former interpretation might
perhaps be accommodated to the limiting case where $\dot{\phi}=\pm
1$, $V(\phi)=0$, the latter view appear to be quite more general
and convenient.

In the general case where $V(\phi)\neq 0$, obviously the evolution
of the universe cannot be dissociated from $V(\phi)$. Thus, the
general and limiting models discussed in this paper can all be
tested by following the evolution of the energy density and
pressure that is predicted by specifying $V(\phi)$ according to
Eqs. (3.11) and (312) for the general case, or to limiting
expressions such as e.g. those in Eqs. (3.14). Determining the
potential $V(\phi)$ this way will allow us to check, moreover, the
extend at which the models with $\dot{\phi}\neq\pm 1$ of Sec. II
share some characteristics of tracking quintessence fields. It
remains an open question, however, to see whether or not the
general model suggested in this paper might add to a possible
solution to the coincidence time problem [19].

It is finally noted that, even though a model involving a
modification of the relativistic point particle quantum
energy-momentum relation and its conversion into a classical field
theory relationship looks more complicate than other models for
dark energy, it however contains some of such simpler models as
limiting cases and shows a much wider applicability.

\acknowledgements

\noindent The author thanks Carmen L. Sig\"{u}enza for useful
discussions. This work was supported by DGICYT under Research
Project BMF2002-03758.

\end{document}